# Origin of the ‘He/N’ and ‘Fe II’ Spectral Classes of Novae

**Robert Williams**

Space Telescope Science Institute, 3700 San Martin Drive, Baltimore, MD 21218

E-mail: wms@stsci.edu

**Abstract**. The spectra of postoutburst novae display either He+N or Fe II lines as the most prominent non-Balmer lines at maximum light. Spectral diagnostics indicate physical conditions for ‘He/N’ spectra that are consistent with their origin in the white dwarf (WD) ejecta, whereas ‘Fe II’ spectra point to their formation in a large circumbinary envelope of gas whose origin is the secondary star. A determining parameter for which of the two types of spectra predominates may be the binary mass ratio $M_{sec}/M_{WD}$. The increasing fraction of novae that are observed to be ‘hybrid’ objects, where both classes of spectra appear sequentially, is explained by changing parameters in the two emitting regions during the postoutburst decline. We argue that most novae may be hybrids that show both types of spectra during decline. The emission line intensity ratio O I λ8446/λ7773 is suggested as a good density diagnostic for the ejecta, and a finding list of emission lines identified in recent spectroscopic surveys of novae is presented as an aid to future line identification work.

## 1. Introduction

The geometry of postoutburst novae involves at least four independent components of the binary systems: the primary white dwarf, the secondary star, the ejecta, and remnants of the possibly re-establishing accretion disk. In addition, strong WD magnetic fields are capable of disrupting the accretion disks, creating accretion funnels that radiate. The interpretation of the spectrum is complicated by the fact that spatial resolution of close mass transfer binaries has not yet been achieved, so uncertainty exists over the relative contributions of the different emitting components. The WD primary stars are depleted in both H and He, and the secondary stars are of late spectral type that range from main sequence to giants that may be evolved and partially stripped stars.

In the aftermath of the discovery and follow up study of supernova SN1987A in the Large Magellanic Cloud Cerro Tololo Inter-American Observatory undertook a spectroscopic survey of galactic novae in the years 1987-93. Approximately 30 postoutburst classical novae were observed with moderate spectral resolutions of 8-16 Å until they entered the nebular phase. The primary aim was to characterize the evolution of novae spectra with broad wavelength coverage and good signal-to-noise data using digital detectors.

A synopsis of the CTIO survey data was presented (Williams et al. 1991, 1994) together with a classification scheme that was proposed for the spectra (Williams 1992). The primary result was that immediately following maximum light all novae spectra fit one of two templates in which the more prominent non-Balmer lines are either He and N lines or Fe II lines. The individual spectra that belong to each class are rather similar to each other and the two templates are designated as ‘He/N’ or ‘Fe II’ class. The two spectral classes differ from each other in terms of their more prominent transitions, their line profiles, velocity structure, ionization level, and time rate of change. An extensive listing of the transitions that have been identified in recent novae surveys and which are associated with the two spectral classes and final nebular spectrum is given in the Appendix.

Statistics of novae in our Galaxy and continuing surveys of M31 novae (Shafter et al. 2011) show that roughly 85% of novae have the ‘Fe II’ type spectrum from maximum light until development of the final emission-line nebular spectrum, while the remaining 15% belong to the ‘He/N’ class. The original CTIO survey showed one unusual nova, LMC 1988 #2, that was unique in that it sequentially exhibited both classes of spectra. In a matter of weeks it evolved from its initial Fe II spectrum to a He/N spectrum before developing its ultimate nebular spectrum. This unique object was classified as a ‘hybrid’ nova based on its unusual presentation of both types of spectra.

## 2. He/N & Fe II spectral classes

The two novae spectral classes possess distinctly different characteristics. ‘He/N’ novae exhibit higher ionization, very broad lines due to high expansion velocities, rectangular profiles, few if any absorption features, and a visible luminosity that decreases relatively rapidly with time. Most of the strongest emission lines are recombination and resonance fluorescence excited lines of H, He, and N. The prominence of He and N transitions, the rectangular line profiles, and the large line widths are suggestive of an origin of this emission in gas that undergoes high velocity, and possibly episodic, ejection from the white dwarf.

‘Fe II’ novae have a spectrum that is marked by numerous low excitation Fe II lines that are collisionally excited and CNO lines in the far red that are excited by recombination and fluorescence scattering. They also have a rich combination of heavy element transitions having a low level of ionization with narrower, more rounded profiles that frequently exhibit prominent P Cygni absorption features characteristic of an optically thick expanding gas. The intensities of Fe II spectra decrease relatively slowly and occasionally even turn around and increase with time during ‘secondary maxima’ episodes weeks and months after the initial outburst. Fe II-type novae also frequently exhibit narrow absorption lines from Fe-peak elements during an interval in the early decline period. The strength of the heavy elements, the smaller line widths that are similar to the escape velocities of late-type stars, and the P Cygni profiles typical of a stellar wind are strongly suggestive of an origin for the Fe II spectra in an envelope of gas with solar-like abundances, as would be expected from the secondary star rather than the WD. This gas may not have been mixed with the outburst ejecta, might be somewhat enriched in heavy elements due to the evolutionary history of the secondary star, but does not show clear signatures of having resided on the WD.

A small fraction of novae have been observed to be the ‘hybrid’ objects that sequentially exhibit both spectral classes. With recent increased observational coverage of novae hybrids are now being found to be more common than previously realized.

## 3. Spectral diagnostics

Various features of nova spectra provide information on physical conditions in the emitting gas. A few salient ones which we believe reveal information on the emitting regions are:

### *3.1. Secondary maxima*

Most novae decline monotonically in visual brightness following their initial peak brightness. A small fraction of them reverse the decline and achieve secondary peaks in brightness that approach the initial visual luminosity within 1-2 magnitudes. Novae that form substantial amounts of dust also experience declines, or dips, in visual brightness that slowly reverse so the light curve reverts to its original rate of decline. The morphology of novae light curves with dust troughs is quite different from the light curves of novae that have secondary peaks and multiple maxima in brightness, which enables one to easily distinguish between the two different phenomena.

Very good examples of the light curves of multiple maxima novae are shown in Strope, Schaefer, & Henden (2010) and Kato & Hachisu (2011, Fig. 6). There has been only sporadic spectral coverage of these secondary maxima events, which last of order a few weeks, until Tanaka et al.'s (2011) thorough spectral study of the nova V5558 Sgr/2007 over a series of such multiple maxima.

The Tanaka et al. data, displayed clearly in Figs. 4, 5, & 6 of their paper, show that during the complete sequence of secondary peaks the optical spectra remain rather constant and unchanged as 'Fe II' spectra, with no evidence of any variation in excitation conditions. At the time of maximum brightness the Balmer and Fe II P Cygni profiles do show more pronounced absorption strengths, indicative of an increasing density and mass loss rate for the putative wind. This behavior during the secondary maxima demonstrates that the increased brightness is due not to a change in effective temperature but to an increased photospheric radius that is caused by a variable wind.

The size of the effective photosphere at the peak of the maxima can be determined assuming (1) an effective temperature of T~7,500 K from the ionization level of the Fe II spectrum, and (2) a luminosity that is approximately 10% of the Eddington luminosity for a 1 $M_{\odot}$ star since many novae have been determined to have $L_{bol} \sim L_{edd}$ for periods of weeks following maximum light (Gallagher & Code 1974; Ney & Hatfield 1978; Gehrz et al. 1980). These constraints dictate a well known fact for novae: the outflowing envelope photospheric radius $R_{phot} \sim 10^2\ R_{\odot}$ is much larger than the binary separation and can be variable.

### *3.2. O I λ8446 & λ7773 relative intensities*

The data from the CTIO and FEROS spectroscopic surveys show correlations between the O I λ8446 and O I λ7773 emission lines that differ between the He/N and the Fe II spectra. The λ8446 line is very strong near maximum light in virtually all novae due to the strong resonance scattering wavelength coincidence between H I Ly-β and O I λ1027, which converts much of the H I recombination energy into O I emission (Strittmatter et al. 1977). No other O I lines are expected to be nearly as strong as λ8446 except those involved in the downward cascade sequence back to the ground state. O I λ7773 is a quintet transition, not a triplet as is λ8446, so its strength should normally be set independently by collisional excitation and recombination, and it should not have an intensity nearly as strong as that of the fluorescence excited λ8446. However, it does appear with strength in many novae and in a few objects has an intensity that even exceeds that of λ8446.

The only realistic possibility for strong excitation of the λ7773 multiplet so its intensity is comparable to that of λ8446 is through collisional de-excitation of the 3p $^3P$ upper level of the λ8446 transition to the 3p $^5P$ upper level of λ7773. This situation has been treated in detail by Bhatia & Kastner (1995) and Kastner & Bhatia (1995), who show that for the low kinetic temperatures, T~7,000 K, that typify early decline spectra and the very large Ly-β intensities required to explain the large emission equivalent widths of O I λ8446 in novae, viz., which require high photoexcitation rates to the O I 3p $^3P$ level, electron densities of order $n_e \sim 10^{12}$ cm$^{-3}$ and greater are needed to produce comparable intensities of λ8446 and λ7773 that also exceed forbidden λ5577. These densities are higher than those of the solar chromosphere so they are almost necessarily associated with stellar photospheric conditions, probably in the form of a dense wind that is almost certainly circumbinary because of the high luminosities involved.

We have measured the observed intensity ratio R ≡ F(λ8446)/F(λ7773) from some of our original low resolution CTIO and more recent high resolution FEROS spectra for which this ratio was observed, and they are listed in Table 1 for those survey objects together with the spectral class of the nova. The R values are approximate because of strong line blending of λ8446 with the Ca II λ8498 IR triplet member, especially in the lower resolution spectra. Also, both λ8446 and λ7773 have appreciable P Cygni absorption components that modify the relative intensities and our measured values of R do take this absorption component into account since it is a legitimate component of the line scattering.

Table 1 indicates a difference in the intensity ratios observed for He/N vs. Fe II novae. Although λ8446 appears in early decline in almost all novae, λ7773 is occasionally not detected in He/N novae. Because of the broad lines and strong continua of He/N novae the lower limits for R in the cases of non-detection of λ7773 are poorly determined but they do indicate that He/N spectra may be formed in lower density gas. On the other hand, some Fe II novae have intensity ratios R<2, with a few being less than unity, which requires formation of Fe II spectra in quite high density gas.

Table 1 lists multiple values of R for various novae for which the line ratio was observed at different times. The subscript entries give the number of days past maximum light or discovery. The flux ratio R generally increases with time, indicating a progressive decrease in the effective photospheric density. However, there are cases where the intensity ratio remains roughly constant for periods of weeks and where it reverses and decreases months after outburst, e.g., for V2574 Oph/04, signaling an increase in the density. Fe II class spectra frequently show little change over months time even though their line profiles signify formation in a gas expanding of order ~$10^3$ km/s. Hence, Fe II class spectra must be formed in dense winds that persist for months after outburst.

The O I λ8446/λ7773 intensity ratio is potentially a good density diagnostic for the gas that forms the spectrum during decline. Not surprisingly, in some objects there appears to be a correlation between the values of the λ8446/λ7773 intensity ratio and the initial appearance of forbidden lines in the spectra.

### *3.3. Hybrid novae*

At maximum visible light classical novae are observed to have either a He/N or a Fe II type spectrum and they normally retain that characteristic spectrum until the expanding gas produces the lower densities and optical depths that lead to the emergence of the nebular emission lines. In the past few decades some novae have been observed to be 'hybrid' objects that transition between the He/N and Fe II spectral types in the weeks following outburst. The fraction of classical novae that have been observed to evolve from one spectral class to the other had been small, roughly 5% of observed novae. However, with increasing spectroscopic follow up of novae more such objects are now being identified.

In addition to the hybrid Nova LMC1988 #2 mentioned earlier Poggiani (2008) observed nova V458 Vul/2007 to be a hybrid object, and in their extensive spectroscopic survey of M 31 novae Shafter et al. (2011) noted a hybrid nova, M31N 2006-10b. All three of these novae were observed as Fe II class at maximum light, after which they transitioned to a He/N spectrum a few weeks following outburst, and which remained until the nebular spectrum developed. A recent outstanding example of a hybrid nova has come from the well studied 2011 outburst of the recurrent novae T Pyxidis, which exhibited a hybrid evolution in having a definite He/N spectrum within a few days of its outburst that then evolved to a clear Fe II spectrum as it continued brightening to maximum light four weeks later (Shore et al. 2011; Ederoclite et al. 2012). It retained its Fe II type spectrum for several weeks and then transitioned back to its original He/N spectrum as it gradually faded and evolved to its emission-line nebular spectrum.

Most novae that are observed to be hybrids have evolved from the Fe II to the He/N class. Evolution the other way, i.e., from He/N to Fe II as exhibited by T Pyx, has rarely been seen. The great majority of novae that are observed initially at maximum light to have He/N spectra retain that spectral type until the emission lines emerge, i.e., they are not hybrids. The fact that the initial He/N phase of the T Pyx spectrum was observed within one day of the outburst and well before maximum light may indicate that hybrid evolution from the He/N to the Fe II class may require the initial spectra to be obtained within hours of the outburst

Based on the hybrid novae observed so far it is evident that spectral evolution between classes can proceed either way, i.e., the spectrum can appear initially as either a He/N or Fe II spectral class and then transition to the other spectral type in a matter of weeks, normally retaining that spectrum until the nebular spectrum emerges. The fact that the two spectral classes, He/N & Fe II, are so different in their basic characteristics indicates that they may originate in different components of gas. The alternative idea cannot be dismissed, however, that the different spectral classes are due to a change in excitation conditions in the ejecta, in which case separate emission components may not be required. We return to this point later.

Hybrid novae transition between the He/N and Fe II classes of spectra because distinct physical conditions predominate at different epochs. Such evolution is probably a common trait of novae but may not normally be observed because of the timing and duration of the different phases. Observational experience indicates that one of the two classes usually remains dominant throughout the entire post-maximum decline so the other component is not detected. Changes in the spectra are most likely energetically driven by continued nuclear burning on the WD surface, however the secondary star might respond to the outburst via pulsations or internal or magnetic activity that could drive some of the spectral changes.

## 4. Current spectral paradigm

A self consistent picture of the spectral evolution of novae has developed over the past 20 years based on modeling the spectra. Its basic tenet is that the optical and UV emission occurs in gas that originates in the outer layers of the white dwarf and has been ejected by it, initially in the form of a shell and subsequently in the form of a wind that continues for months to years. Hauschildt and collaborators' detailed work with the PHOENIX radiative transfer atmosphere code (Hauschildt et al. 1995, 1996, 1997) showed that they could fit the spectra of novae from the time of the early fireball to months after maximum light with a wind having a power-law density distribution. One of their significant findings was the demonstration that non-LTE effects are very important in treating line opacities which in a rapidly expanding medium mimic continuum absorption, and that these non-LTE effects can be more important than abundances in determining the general nature of the spectrum (Hauschildt et al. 1996).

An important result of the Hauschildt PHOENIX models is that they support mass ejection and a wind that can have solar element abundances because the line features are often more sensitive to non-LTE effects and the assumed density distribution than to abundances (Hauschildt et al. 1997). This was demonstrated in the analysis of nova OS And/1986 by Schwarz et al. (1997), who modeled its 'Fe II' spectrum successfully with solar abundances.

Kato & Hachisu (2009, 2011) have conducted studies of the photospheric regions of novae in which they consider the hydrodynamics of the emitting gas as it transitions from a static to an optically thick wind geometry. They find they can fit the light curves of novae well, including those that have multiple peaks in brightness, by models in which a WD wind undergoes variations that are triggered by the companion star energizing the wind from frictional drag as it orbits within a common envelope. They consider the wind to originate on the WD. Similarly, the results of independent investigations by other groups have assumed the white dwarf to be the source of mass loss and the emitting gas that produces the spectrum of the postoutburst nova (Bode 2010), except in cases where the secondary star is a giant (Anupama & Mikolajewska 1999). As a result of these investigations there has been general consensus that the dominant source of the postoutburst spectrum is gas ejected from the WD (Shore et al. 2011).

## 5. Competing spectrum components

The central question governing the spectral evolution of novae remains: are the He/N and Fe II spectra formed in the same ejecta or do they originate in distinctly different components of gas? The arguments presented in *§.III* favor the latter:

1. The Fe II and He/N line profiles are different in both shape and velocity width. A general result of winds and stellar ejecta is that coupling of collisional and radiative processes with hydrodynamical constraints typically causes terminal velocities to be within factors of 2-3 of the escape velocity from the object. The smaller Fe II absorption velocities are consistent with their association with the secondary star or circumbinary envelope whereas the higher velocities of the He/N spectra are consistent with ejection from the WD.
2. The more solar-like abundances allowed by analyses of the Fe II spectra and especially their higher densities inferred from the O I λ8446/λ7773 intensity ratios indicate a significant departure from the lower density and He- and N-enriched gas of He/N spectra.

The extent to which these arguments are conclusive depends on the other parameters involved in the formation of spectra but they do give credibility to the idea that both the WD and the secondary star contribute separately to the spectra.

If one accepts the supposition that two independent gaseous components do contribute to postoutburst novae emission a natural sequence of spectral evolution follows from the outburst. The outburst is triggered by accretion onto the WD which produces a thermonuclear runaway in partially degenerate gas. Gas that is depleted in H due to the TNR outburst proton-capture reactions, i.e., is enhanced in He and N (Starrfield, Truran, & Sparks 2000; Yaron et al. 2005), is ejected by the outburst at high velocities that exceed the WD escape velocity. *These ejecta produce the He/N spectrum and should be the initial dominant spectrum of every nova.* Depending on the mass ejected the He/N lines could be optically thin or thick, and the ejecta will impact the secondary star less than one hour after outburst.

Depending on the proximity of the secondary star, the WD ejecta mass, and the radiation field the secondary star should be stimulated by the TNR to eject mass. Aspects of this situation have been studied by Campbell et al. (2011) using the SPH ‘splash’ code, and by Drake & Orlando (2010) using the FLASH hydrodynamic code and although they do not model the response of the secondary star to the ejecta blast wave they do show effects of the ejecta colliding with the secondary. Similarly, hydro calculations of Sytov et al. (2007, 2009) have demonstrated that mass loss can occur through both the L1 and L3 points and take the form of a circumbinary envelope from the bow shock created by orbital motion of the WD and accretion disk, and this material, much of it pre-outburst, will not have resided on the WD.

The excitation of this gas will give rise to the Fe II-type spectrum, whose abundances should be more solar-like, i.e., characteristic of an evolved secondary, and with lower expansion velocities characteristic of the escape velocities of the secondary star and circumbinary envelope. Thus, shortly after outburst there should generally be two distinct components of emitting and absorbing gas.

We assume that at least $10^{-4}$ of the total energy $E_{TNR}$~$10^{45}$ erg created by the TNR is emitted in the form of γ-rays in the first few hours of the outburst (Starrfield 1999), which are absorbed by the outer layers of the secondary star to a depth in column density determined by the Thomson scattering cross section, i.e., $10^{25}$ $cm^{-2}$. The thermal energy of the absorbing gas in the secondary within the above column density is of order $E_{therm} \sim 10^{37}$ erg for temperatures $T \sim 10^4$ K. Thus, assuming that a fraction of the TNR $\gamma$-rays escape outward unabsorbed by the outer layers and ejecta of the WD, the absorption of $10^{41}$ erg by the secondary outer layers must lead to an expansion of the secondary star beyond its Roche lobe. The amount of mass affected is only of order $10^{-10}$ $M_{\odot}$ but the injection of such energy should stimulate greater mass loss.

The TNR on the WD surface provides the energy for the outburst so an important element in validating the above hypothesis of separate emission components is understanding how the circumbinary envelope from the secondary star is created. Assuming that the initial source of energy must come from the WD a key parameter in activating the secondary must be the angular size it subtends at the WD, which depends largely on the mass ratio $q = M_{sec}/M_{WD}$ as long as the secondary fills its Roche lobe. Thus, the spectral evolution of novae could depend as much or more on the binary mass ratio, q, as on the individual stellar masses, element abundances, orbital period, and mass transfer rate. In particular, the spectra of novae with q<<1, where the secondary star subtends small angular size[1], would be expected to be dominated by the He/N spectrum of the WD ejecta rather than the Fe II spectrum. Objects with q<<1 probably tend to have relatively high WD masses, for which one would also expect short outburst recurrence times and rapid decline rates for these novae---and for which He/N spectra are, in fact, usually observed.

The initial dominance of the high velocity He/N emission may last only a few hours depending upon the extent of the mass loss from the secondary star, or it could remain dominant during the entire post-maximum decline. The alternative, depending on the response of the secondary star to the outburst, is initiation of a phase of enhanced secondary star mass loss whose line and continuum spectrum will predominate over the initial WD He/N component. This changeover in the predominance of the two gas components could be the cause of the oft observed 'pre-maximum halt' that precedes maximum light. The Fe II spectrum would then remain dominant, as is observed for the large majority of novae, if the mass loss rate of the wind from the secondary remains sufficiently high to predominate over emission from the WD ejecta. In this case, the large majority of novae are expected to be 'hybrid' objects in terms of their spectral development.

The dominant intensity of the 'Fe II' spectrum is due to the large photospheric radius in the wind which produces a stronger continuum flux than the peak intensity of the He/N emission lines. The He/N gas is ejected first and has higher velocities so it exists outside the Fe II emitting component and, in principle, could still be detectable. The fact that 85% of novae radiate an Fe II-type spectrum means that it normally dominates as the expanding gas decreases in density and enters the nebular emission-line phase. This requires that the secondary star respond to the outburst with a wind and substantial continuing mass loss that exceeds the mass ejected from the WD.

The possibility that the binary mass ratio or separation may be important in determining which type of spectrum predominates in a nova during the postoutburst decline is testable. Stellar mass ratios for various eclipsing novae systems have been determined and are listed in the Ritter & Kolb (2003) catalogue. The majority of them occur in the q>0.5 domain. Only one object having a low value of q<0.25 has good spectral information, V603 Aql/1918, and its spectral class remains ambiguous from the published records.

An alternative way of looking for the effect of binary mass ratio on spectral class is to see if the spectral classes of novae depend on their orbital period. Most novae with periods within or below the 2-3$^{h}$ period gap are believed to have massive WDs that result in low q-values for these systems. We have therefore searched those novae in the Ritter & Kolb (2003) catalogue that have orbital periods P<3 hr together with published spectra since 1990, when serious digital spectra follow up of novae commenced. The relevant short orbital period novae, listed with their spectral class, are: V2362 Cyg/2006 (Fe II), V1974 Cyg/1992 (hybrid: Fe II > He/N), KT Eri/2009 (He/N), V2491 Cyg/2008 (He/N), V597 Pup/2007 (Fe II), V351 Pup/1991 (Fe II), and V5116 Sgr/2005#2 (Fe II). This admittedly small number of novae hardly constitutes a meaningful sample and it is at best only marginally suggestive that He/N spectra might be more common among the short period, and therefore ostensibly low $M_{sec}/M_{WD}$, novae.

---

[1] Using the relationship derived by Paczynski (1971) for the radius of a Roche-lobe filling secondary star it follows that the solid angle subtended by the secondary from the WD varies from 1.6 ster for q=1 to 0.3 ster for q=½.

A class of recurrent novae exist with giant secondary stars whose orbital periods are long. Their prototypes are RS Oph and T CrB, and they include V745 Sco and V3890 Sgr (Schaefer 2010). All four RNe have M giant secondaries and orbital periods P>200 days. Given the relatively large distance of the secondary stars from the WD it is reasonable to expect that the outburst may have less effect in dictating the resulting early spectrum for these RNe. According to the paradigm suggested here this would make an Fe II spectrum less likely. The large extent of the M giant low density atmospheres provide an ideal place for emission lines to form and, in fact, the early decline spectra of all the above RNe do show strong, narrow emission lines superposed on a broader spectrum of ejecta. The helium lines are the stronger non-Balmer lines, which is what one would expect in this situation where the ejecta predominate, although the very extended atmospheres of the secondaries do make a definite contribution to the spectra.

Finally, the survey of novae in M33 by Shafter et al. (2012) found a preponderance of He/N and Fe IIb novae in that galaxy compared with the normal 75% of Fe II novae that occur in the Galaxy and M31. Shafter and colleagues attribute this difference to a possibly younger population of novae that contain more massive WDs. If true this explanation would be consistent with the paradigm suggested here in that one would expect the He/N-rich WD ejecta to dominate the spectrum in novae with massive WDs. The alternative possibility that different abundances may be responsible for the different nova populations may be operative but it is not obvious why abundances should affect the He/N vs. Fe II distribution.

**6. Test of the hypothesis**

The most straightforward way to establish separate emitting components is to detect radial velocity variations for the He/N and Fe II spectra, especially in hybrid objects where both spectral components are observed since they should have different velocity phases. This may be possible with high spectral resolution, high S/N observations of highly inclined systems. The difficulty may be that the emission from either an extended circumbinary envelope or the WD ejecta is almost certainly a mixture of gas that has been ejected over multiple binary cycles and therefore radial velocity signatures from the stellar orbital motions may be averaged out.

Other evidence for separate Fe II and He/N emission regions may come from transient 'thea' absorbing systems that were studied in the FEROS survey (Williams et al. 2008). *Thea* absorption has so far been observed only in Fe II-type spectra and Williams & Mason (2010) argued that the absorbing gas is present in pre-outburst shells. However, high resolution observations of *thea* absorption in the recent T Pyxidis outburst by Ederoclite et al. (2012) detected a *thea* system that was not present at the time of outburst but emerged only some weeks afterward when T Pyx had entered its Fe II spectral phase, casting doubt on the Williams & Mason hypothesis. The sudden appearance of *thea* absorption with the Fe II spectrum in T Pyxidis suggests an association between them, i.e., creation of *thea* absorption by the Fe II component of gas, whose abundances are similar to those of the secondary star. If *thea* systems continue to be associated exclusively with Fe II spectra rather than He/N spectra it would support the existence of separate emission components.

## 7. Summary

The nova outburst ejects both He- and CNO-enriched gas from the WD surface at velocities typically in excess of ~2,500 km/s, and these ejecta are the source of the 'He/N' spectrum. The ejecta impact the secondary star and together with the heating of the outer layers of the secondary star by radiation from the outburst they initiate an increase in mass loss by the secondary. Some of this gas may be accreted onto the WD but a significant amount is lost via a wind from the secondary. Some of the mass loss from the secondary star will eventually re-establish the accretion disk but most of it will become circumbinary. Depending on the secondary mass loss rate and wind geometry, which could vary over a matter of hours to weeks, this material may be insubstantial---in which case the dominant spectrum remains of the He/N type. More commonly, the secondary contributes to a substantial expanding photosphere that then becomes the major source of the Fe II spectrum. As the wind decreases the combined ejecta become optically thin and the spectrum progressively evolves into the nebular emission-line phase, which consists of a mixture of both components of gas from the two stars. Thus, abundance analyses of novae based on emission lines in the nebular phase must take into account the fact that the gas represents a mix of the H-depleted, CNO-enhanced ejecta from the WD and the possibly evolved and stripped secondary star.

The fact that hybrid novae virtually always evolve only from the broad Fe IIb class, not the narrow Fe IIn class, to the He/N spectral class is notable. Why do narrower line Fe II novae not evolve so that the He/N ejecta are eventually revealed? The likely reason is that the higher expansion velocities of Fe IIb novae lead to a much more rapid decrease in density that drives the optical depth of the Fe II emitting gas lower to reveal the He/N ejecta emission. In addition, the narrow line Fe II novae may also have longer continuing mass loss from the secondary star than the broad line Fe IIb objects.

Previous studies of novae have acknowledged the secondary star as a contributor to the formation of the spectrum but few have advocated that direct mass loss from it is the major source of the emission. The disparate natures of the He/N and the Fe II spectra are strong arguments for their separate formation in different sources, i.e., the WD and the secondary star. Continued study of postoutburst novae should soon provide more concrete evidence for the origins of the postoutburst radiation, with postoutburst radial velocity studies of high inclination systems most likely to provide the best indication of the origins of the different classes of spectra.

**Table 1**
**Intensity Ratio O I λ8446/λ7773 in Novae**

| Nova | Spectral Class | F(λ8446/λ7773) | Nova | Spectral Class | F(λ8446/λ7773) |
|---|---|---|---|---|---|
| V2214 Oph/88 | Fe II | $47_{35}$ | V1187 Sco/04 #2 | Fe II | $7.8_{9}$, $14_{12}$, $14_{34}$ |
| LMC 1988 #1 | Fe II | $1.1_{2}$, $6.1_{29}$ | V2574 Oph/04 | Fe II | $12_{28}$, $13_{122}$, $15_{128}$ |
| V443 Sct/89 | Fe II | $1.8_{15}$, $2.6_{210}$ | V5114 Sgr/04 | Fe II | $0.50_{1}$, $0.50_{2}$, $2.6_{9}$, $11_{23}$, $19_{32}$ |
| V977 Sco/89 | Fe II | $22_{28}$, $34_{46}$ | V378 Ser/05 | Fe II | $1.3_{15}$, $2.2_{39}$, $21_{87}$ |
| V3890 Sgr/90 | Hybrid | $17_{28}$, $20_{34}$, $44_{42}$ | V382 Nor/05 | Fe II | $5.2_{6}$, $19_{17}$, $37_{41}$, $34_{72}$ |
| V868 Cen/91 | Fe II | $1.3_{3}$, $2.5_{25}$, $4.0_{38}$, $7.9_{73}$, $4.2_{94}$,$3.4_{282}$ | V1663 Aql/05 | Fe II | $11_{28}$, $14_{95}$, $10_{108}$ |
| LMC 1991 | Fe II | $1.8_{-4}$, $0.49_{0}$, $9.4_{13}$ | V476 Sct/05 #1 | Fe II | $4.6_{15}$, $19_{38}$ |
| V2264 Oph/91 | Fe II | $2.6:_{13}$, $3.5_{18}$, $7.9_{31}$, $17_{66}$, $14_{87}$ | V5116 Sgr/05 #2 | Fe II | $0.9_{3}$, $12_{22}$, $33_{60}$ |
| V351 Pup/91 | Fe II | $5.0_{11}$, $29_{32}$, $27_{40}$ | LMC 2005 | Fe II | $1.3_{6}$, $1.8_{45}$, $15_{91}$ |
| V4169 Sgr/92 | Fe II | $6.9_{20}$, $15_{88}$ | V2575 Oph/06 | Fe II | $4.7_{37}$ |
| V992 Sco/92 | Fe II | $1.7_{64}$, $3.1_{93}$ | ------ | --- | ----- |
| LMC 1992 | Fe II | $2.8_{15}$, $4.5_{20}$, $16_{77}$ | V745 Sco/89 | He/N | $>65_{23}$ |
| V705 Cas/93 | Fe II | $23_{68}$ | LMC 1990 #1 | He/N | $>180_{4}$ |
| V2295 Oph/93 | Fe II | $9.4_{14}$ | V838 Her/91 | He/N | $>50:_{9}$ |
| V2573 Oph/03 | Fe II | $5.6_{22}$, $3.0_{50}$, $5.9_{72}$ | V382 Vel/99 | He/N | $2.5_{5}$, $3.6_{7}$, $7.6_{14}$, $19_{69}$ |
| V1186 Sco/04 #1 | Fe II | $5.8_{31}$ | V5115 Sgr/05 #1 | He/N | $3.4_{6}$, $21_{30}$ |

## APPENDIX

SPECTROSCOPIC FINDING LIST FOR EMISSION LINES IN NOVAE

The identification of lines in novae spectra is a difficult task because irregular profiles, large line widths, and variable velocity shifts make the true wavelengths of features difficult to establish. Unusual excitation conditions and abundances in complex binary systems also make an important determiner of correct IDs, ‘astrophysical reasonableness’, difficult to apply. The inevitable blending of transitions and multiplets can be difficult to untangle although following the change in time of the spectrum does help reveal line blends. High spectral resolution and good signal-to-noise are important for reliable line IDs, and the construction of models using detailed radiative transfer codes has helped to establish a list of transitions that are being used to study the nova phenomenon.

Table A1 lists the different transitions that have been identified in postoutburst novae in recent years, and they are categorized by the two early decline nova spectral classes, Fe II and He/N, in which they are primarily observed. Some of the multiplet numbers are provided in parentheses.

This list is work in progress in the sense that uncertainties in line identifications have caused us to list possible candidate lines that may actually be due to nearby transitions. This list therefore errs on the side of inclusion. With time and increasing use of higher spectral resolution the list of spectral features will change and become more reliable. The author will continue to refine this list and make the updated list available to the community.

## Table A1
**FINDING LIST FOR OPTCAL EMISSION LINES IN CLASSICAL NOVAE**

| Line ID | Spectral Class *'Fe II'* | *'He/N'* | *'Nebular'* |
|---|---|---|---|
| He II  3203 | | | X |
| Fe II  3228 (6) | X | | |
| Fe II  3256 (1) | X | | |
| Fe II  3277 (1) | X | | |
| Fe II  3281 (1) | X | | |
| [Ne III]  3343 | | | X |
| [Ne V]  3346 | | | X |
| [Ne V]  3426 | | | X |
| O VI  3435 | | | X |
| O III  3444 | | | X |
| [Fe VI]  3492 | | | X |
| [Fe VI]  3556 | | | X |
| [Fe VII]  3586 | | | X |
| [Fe VI]  3663 | | | X |
| O III  3713 (14) | | | X |
| [S III]  3722 | | | X |
| [O II]  3727 | | | X |
| H I  3734 | X | X | X |
| Ca II  3706/37 (3) | X | | |
| H I  3750 | X | X | X |
| [Fe VII]  3759 | | | X |
| H I  3771 | X | X | X |
| H I  3798 | X | X | X |
| O VI  3811 | | | X |
| [Fe VI]  3814 | | | X |
| O VI  3834 | | | X |
| H I  3835 | X | X | X |
| Mg I  3835 (3) | X | | |
| Mg II  3849 (5) | X | | |
| He II  3858 | | | X |
| Si II  3858 (1) | X | | |
| [Ne III]  3869 | | | X |
| H I  3889 | X | X | X |
| He I  3889 | | X | X |
| Al II  3901 (1) | X | | |
| He II  3923 | | X | |
| Ca II  3934 (K) | X | | |
| Ca II  3968 (H) | X | | |
| [Ne III]  3968 | | | X |
| H I  3970 | X | X | X |
| [Fe XI]  3987 | | | X |
| He I  4026 | | X | X |
| [Fe V]  4026 | | | X |
| [S II]  4069 | | | X |
| [Fe V]  4071 | | | X |
| [S II]  4076 | | | X |
| H I  4102 | X | X | X |
| He I  4144 | | X | X |
| Fe II  4179 (28) | X | | |
| [Fe V]  4181 | | | X |
| C III  4187 | | X | |
| He II  4200 | | X | X |
| [Fe V]  4229 | | | X |
| Fe II  4233 (27) | X | | |
| [Fe II]  4244 | | | X |
| C II  4267 | | X | |
| [Fe II]  4287 | | | X |
| Fe II  4297 (28) | X | | |
| Fe II  4303 (27) | X | | |
| H I  4340 | X | X | X |
| Fe II  4352 (27) | X | | |
| [Fe II]  4359 | | | X |
| [O III]  4363 | | | X |
| Fe II  4385 (27) | X | | |
| He I  4388 | | X | X |
| [Fe II]  4416 | | | X |
| Fe II  4417 (27) | X | | |
| He I  4438 | | X | X |
| [Fe II]  4458 | | | X |

| Line ID | Spectral Class *'Fe II'* | *'He/N'* | *'Nebular'* |
|---|---|---|---|
| He I  4471 | | X | X |
| Mg II  4481 (4) | X | | |
| Fe II  4491 (37) | X | | |
| Fe II  4508 (38) | X | | |
| Fe II  4515 (37) | X | | |
| N III  4517 (3) | | X | X |
| Fe II  4523 (38) | X | | |
| He II  4542 | | X | X |
| Fe II  4549 (38) | X | | |
| Fe II  4556 (37) | X | | |
| Mg I]  4571 | X | | |
| Fe II  4584 (38) | X | | |
| N V  4609 (1) | | X | X |
| Fe II  4629 (37) | X | | |
| N III  4638 | | X | X |
| C IV  4658 | | | X |
| [Fe III]  4658 | | | X |
| Al II  4663 (2) | X | | |
| He II  4686 | | X | X |
| [Fe III]  4702 | | | X |
| He I  4713 | | X | X |
| [Ne IV]  4721 | | | X |
| [Ar IV]  4740 | | | X |
| [Fe III]  4755 | | | X |
| [Fe II]  4815 | | | X |
| H I  4861 | X | X | X |
| [Fe III]  4881 | | | X |
| [Fe VII]  4893 | | | X |
| He I  4922 | | X | X |
| Fe II  4924 (42) | X | | |
| [Ca VII]  4939 | | | X |
| [Fe VII]  4942 | | | X |
| N V  4945 | | X | X |
| [O III]  4959 | | | X |
| [Fe VI]  4967/72 | | | X |
| [Fe VII]  4989 | | | X |
| N II  5001 (24) | | X | |
| [O III]  5007 | | | X |
| He I  5016 | | X | X |
| Fe II 5018 (42) | X | | |
| He I  5048 | | X | X |
| [Fe VI]  5146 | | | X |
| [Fe VII]  5158 | | | X |
| [Fe II]  5159 | | | X |
| Fe II  5169 (42) | X | | |
| [Fe VI]  5176 | | | X |
| Mg I  5178 (2) | X | | |
| Fe II  5198 (49) | X | | |
| [N I]  5199 | | | X |
| Fe II  5235 (49) | X | | |
| [Fe II]  5262 | | | X |
| Fe II  5265 (48) | X | | |
| [Fe III]  5270 | | | X |
| Fe II  5276 (49) | X | | |
| [Fe VII]  5276 | | | X |
| Fe II  5284 (41) | X | | |
| O VI  5292 | | X | X |
| [Fe XIV]  5303 | | | X |
| [Ca V]  5309 | | | X |
| Fe II  5317 (49,48) | X | | |
| Fe II  5363 (48) | X | | |
| He II  5412 | | X | X |
| [Fe VI]  5424 | | | X |
| Fe II  5425 (49,48) | X | | |
| N II  5479 (29) | | X | |
| Mg I  5528 (9) | X | | |
| [Ar X]  5533 | | | X |
| Fe II  5535 (55) | X | | |
| [O I]  5577 | | | X |

| Line ID | Spectral Class | | |
|---|---|---|---|
| | *'Fe II'* | *'He/N'* | *'Nebular'* |
| [Ca VII]  5619 | | | X |
| [Fe VI]  5631 | | | X |
| [Fe VI]  5677 | | | X |
| N II  5679 (3) | | X | |
| Na I  5686 (6) | X | | |
| Al III  5706 (2) | X | | |
| [Fe VII] 5721 | | | X |
| [N II]  5755 | | | X |
| C IV  5805 (1) | | | X |
| He I  5876 | | X | X |
| Na I  5892 (D) | X | | |
| N II  5938 (28) | | X | |
| Si II  5958/79 (4) | X | | |
| Fe II  5991 (46) | X | | |
| Fe II  6084 (46) | X | | |
| [Fe VII] 6086 | | | X |
| [Ca V]  6087 | | | X |
| Fe II  6148 (74) | X | | |
| Na I  6159 (5) | X | | |
| O VI  6200 | | | X |
| He II  6234 | | | X |
| Al II  6237 (10) | X | | |
| Fe II  6248 (74) | X | | |
| [O I]  6300 | | | X |
| He II  6311 | | X | X |
| [S III]  6312 | | | X |
| N II  6346 (46) | X | | |
| Si II  6347/71 (2) | X | | |
| [O I]  6364 | | | X |
| [Fe X]  6375 | | | X |
| Fe II  6417 (74) | X | | |
| [Ar V]  6435 | | | X |
| Fe II  6456 (74) | X | | |
| N II  6482 (8) | | X | |
| N I  6486 (21) | X | | |
| [N II]  6548 | | | X |
| H I  6563 | X | X | X |
| He II  6560 | | X | |
| [N II]  6584 | | | X |
| He I  6678 | | X | X |
| He II  6683 | | X | |
| [S II]  6716/31 | | | X |
| Al II  6830 (9) | X | | |
| He II  6891 | | X | |
| O I  7002 (21) | X | | |
| [Ar V]  7006 | | | X |
| Al II  7049 (3) | X | | |
| He I  7065 | | X | X |
| [Ar III]  7136 | | | X |
| [Fe II]  7155 | | | X |
| [Ar IV]  7171 | | | X |
| He II  7178 | | X | X |
| C II  7235 (3) | X | X | |
| [Ar IV]  7237 | | | X |
| [Ar IV]  7263 | | | X |
| He I  7281 | | X | X |
| Fe II  7308 (73) | X | | |
| [O II]  7320/30 | | | X |
| N I  7452 (3) | | X | |
| Fe II  7462 (73) | X | | |
| Al II  7471 (21) | X | | |
| O I  7477 (55) | X | | |
| He II  7593 | | X | X |
| N IV  7703 | | X | X |

| Line ID | Spectral Class | | |
|---|---|---|---|
| | *'Fe II'* | *'He/N'* | *'Nebular'* |
| Fe II  7712 (73) | X | | |
| O IV  7713 | | | X |
| [S I]  7725 | | | X |
| C IV  7726 | | | X |
| [Ar III]  7751 | | | X |
| O I  7773 (1) | X | | |
| [P II]  7876 | | | X |
| [Fe XI]  7892 | | | X |
| Mg II  7896 (8) | X | | |
| [Mn IX]  7968 | | | X |
| N I  7904 | X | | |
| N I  8166/8202 | X | | |
| Na I  8191 (4) | X | | |
| N I  8212 (2) | X | | |
| O I  8227 (34) | X | | |
| Mg II  8232 (7) | X | | |
| He II  8237 | | X | X |
| Ca II  8251/03 (13) | X | | |
| C I  8335 (10) | X | | |
| O I  8446 (4) | X | X | X |
| Ca II  8498 (2) | X | | |
| H I  8502 (P16) | X | X | X |
| Ca II  8542 (2) | X | | |
| H I  8545 | X | X | X |
| H I  8598 | X | X | X |
| N I  8617 (8) | X | X | |
| Al II  8641 (4) | X | | |
| Ca II  8662 (2) | X | | |
| H I  8665 | X | X | X |
| N I  8692 (1) | X | X | |
| [C I]  8727 | | | X |
| H I  8750 (P12) | X | X | X |
| Mg I  8807 (7) | X | | |
| H I  8863 (P11) | X | X | X |
| H I  9015 (P10) | X | X | X |
| N I  9029/60 (15) | X | X | |
| [S III]  9069 | | | X |
| C I  9087 (3) | X | | |
| N I  9208/9188 | X | X | |
| H I  9229 (P9) | X | X | X |
| Mg II  9218/44 (1) | X | | |
| O I  9264 (8) | X | | |
| He II  9345 | | X | X |
| N I  9395 (7) | X | X | |
| C I  9406 (9) | X | | |
| [S III]  9531 | | | X |
| He II  9542 | | X | X |
| H I  9546 (P8) | X | X | X |
| C I  9658 (2) | X | | |
| He II  9762 | | X | X |
| N I  9831 (19) | X | | |
| [S VIII]  9911 | | | X |
| He II  10045 | | X | X |
| H I  10049 (P7) | X | X | X |
| Al II  10090 (6) | X | | |
| N I  10117 (18) | X | | |
| He II  10124 | | X | X |
| C I  10124 | X | | |
| [N I]  10404 | | | X |
| N I  10526 (28) | X | X | |
| C I  10541 (20) | X | | |
| C I  10693 (1) | X | | |
| He I  10830 | X | X | X |
| Mg II  10926 (3) | X | | |
| H I  10938 (P6) | X | X | X |